\documentclass[twocolumn,preprintnumbers,amsmath,amssymb]{revtex4-1}
\usepackage{graphicx}
\usepackage{amsmath}
\usepackage{amssymb}
\usepackage{dcolumn}

\begin{document}

\title{Periodic spin textures in a degenerate $F=1$ $^{87}$Rb spinor Bose gas}
\author{M. Vengalattore$^1$}
\thanks{Present address: Laboratory of Atomic and Solid State Physics,
Cornell University, Ithaca, NY 14853.}
\author{J. Guzman$^1$}
\author{S. R. Leslie$^1$}
\author{F. Serwane$^{1}$}
\author{D. M. Stamper-Kurn$^{1,2}$}
\thanks{Electronic address: dmsk@berkeley.edu}
\affiliation{
    $^1$Department of Physics, University of California, Berkeley CA 94720 \\
    $^2$Materials Sciences Division, Lawrence Berkeley National Laboratory, Berkeley, CA 94720}
\date{\today }

\begin{abstract}
We report on the spin textures produced by cooling unmagnetized
$^{87}$Rb $F=1$ spinor gases into the regime of quantum degeneracy.
At low temperatures, magnetized textures form that break
translational symmetry and display short-range periodic magnetic
order characterized by one- or two-dimensional spatial modulations
with wavelengths much smaller than the extent of the
quasi-two-dimensional degenerate gas. Spin textures produced upon
cooling spin mixtures with a non-zero initial magnetic quadrupole
moment also show ferromagnetic order that, at low temperature,
coexists with the spatially modulated structure.
\end{abstract}


\maketitle

Coherent quantum fluids exhibiting spontaneous spatial order have
garnered widespread  attention in connection to possible supersolid
phases of matter \cite{kim04supersolid} and the ground states of
high-T$_c$ superconductors \cite{oren00,hoff02fourunitcell} and
other correlated electronic materials \cite{mill05mang}. Such
intrinsically heterogenous quantum fluids may arise due to the interplay
between multiple order parameters \cite{mill05mang}, the influence
of adjacent ground states with differing tendencies \cite{sach00}
or  the presence of competing interactions \cite{seul95,dago05}.

Recent observations hint at similar phenomenology in a magnetic
quantum gas, the $F=1$ spinor Bose gas of $^{87}$Rb. Early studies
of $^{87}$Rb spinor condensates \cite{schm04,chan04} suggested that
their magnetic properties were governed solely by the spin-dependent
contact interaction. This interaction, with mean-field energy $-
|c_2| n \langle \mathbf{F} \rangle^2$, favors spin states with
maximum magnetization \cite{ho98,ohmi98}; here, $c_2$ is related to
$s$-wave scattering lengths for interatomic collisions, $n$ is the
number density, and $\mathbf{F}$ is the dimensionless vector spin.
Together with the kinetic energy cost for spatial variation of the
superfluid vector order parameter, this local interaction favors a
simple, homogeneous, ferromagnetically ordered spinor condensate.
However, recent works
\cite{yi04dipolar,yi06texture,kawa07observe,veng08helix} point to
the significance of magnetic dipolar interactions in determining
magnetization dynamics in $^{87}$Rb gases with large spatial extent.
This interaction is long-ranged and spatially anisotropic, and, as
in magnetic thin films, may favor inhomogeneous and spatially
ordered spin textures.  How the competition between the spatially
isotropic contact interaction and the anisotropic long-range dipole
interaction resolves itself in such degenerate spinor gases is the
subject of several recent theoretical investigations
\cite{taka07classical,cher09roton,kawa09crystal_note,zhan09vlattice_note,kjal09},
and remains an open experimental question.

Here, we address this question by examining the magnetic order in
gases produced upon  cooling unmagnetized thermal spin mixtures into
the regime of quantum degeneracy.   At their lowest temperatures,
these quantum fluids break translational symmetry to arrive at
magnetized spin textures that, while varying between samples,
consistently display spatial modulations of similar morphology and
length scale independent of initial conditions and equilibration
time.

This work differs from previous studies of spin textures of spinor
Bose condensates \cite{sadl06symm,veng08helix,lesl09amp} (also for
$F=2$ gases \cite{kronjager}) in two crucial respects. First, in
previous works, modulated spin textures arose from dynamical
instabilities of long-range-ordered spin-polarized Bose-Einstein
condensates.  Thus, the different characteristics of the
magnetization structures observed in those works, while providing
insight on the magnetic interactions present in the spinor gas,
derived from the specific unstable initial states chosen.  Indeed,
the magnetization patterns produced through two different dynamic
instabilities, the first being the spin-mixing instability of a
paramagnetic condensate quenched across a phase transition
\cite{sadl06symm,lesl09amp} and the second being an instability of a
helical spin texture \cite{veng08helix}, showed markedly different
spatial correlations. As such, the present study is aimed at
revealing the magnetic phases favored intrinsically by the spinor
gas.  Second, this work presents the first observation of spin
textures in degenerate Bose gases at variable temperature, whereas
previous works examined structure formation only at near-zero
temperature. The addition of a significant normal gas component
should add new means for dissipation and relaxation from
non-equilibrium magnetization states.

\section{Experimental method}

The local one-body density matrix of a spin-1 atomic gas describes
both rank-1 and rank-2 polarization moments, corresponding to the
vector spin and magnetic quadrupole moments of the atoms,
respectively. For our experiments, we prepared non-degenerate
optically trapped gases characterized by homogeneous fractional
populations $(\zeta_1, \zeta_0, \zeta_{-1})$ in the three
eigenstates of $F_z$.  These gases were unmagnetized, i.e.\
characterized by zero vector spin, given the dual constraints of
$\zeta_1 = \zeta_{-1}$ (zero longitudinal magnetization) and the
absence of coherence among the sublevels (zero transverse
magnetization). These constraints still allow for a non-zero rank-2
(magnetic quadrupole) polarization moment, for $\eta = \zeta_0 -
\zeta_1 \neq 0$, which breaks spin-rotational symmetry by favoring
the $\hat{z}$ axis.

To prepare such spin mixtures, we began with non-degenerate,
longitudinally spin-polarized gases trapped at the focus of an
elliptically focused, linearly polarized, 825-nm wavelength laser
beam.  We then produced incoherent spin mixtures by applying
resonant $\pi/2$ rf pulses while also applying a 50 mG/cm magnetic
field gradient to the gas.  Diffusion of the non-degenerate atoms in
the inhomogeneous field eliminated transverse coherences, as
ascertained by probing for Larmor precession in the thermal gas
\cite{higb05larmor}.  Applying a single rf pulse resulted with an
$\eta = 1/4$ spin mixture, while applying the pulse-diffusion
sequence repeatedly yielded a fully unpolarized gas with $\eta = 0$.

Following their preparation, spin mixtures were evaporatively cooled
by gradually lowering the intensity of the optical trapping beam,
typically over 200 ms. During this process, the trap frequencies,
with values of $(\omega_x, \omega_y, \omega_z) = 2 \pi (84, 1000,
10) \, \mbox{s}^{-1}$ at the initial trap depth of $U/k_B \simeq 10
\, \mu\mbox{K}$, decreased as $\omega_{x,y,z} \propto \sqrt{U}$ and
the gas temperature was found to scale as $k_B T = 0.11 \, (U/k_B -
0.8\, \mu\mbox{K})$, with all Zeeman sublevels attaining the same
temperature.  The offset in the formula for the temperature $T$
accounts for the effects of gravity in the $\hat{y}$ direction.

\begin{figure}[tb]
\centering
\includegraphics[width=0.45\textwidth]{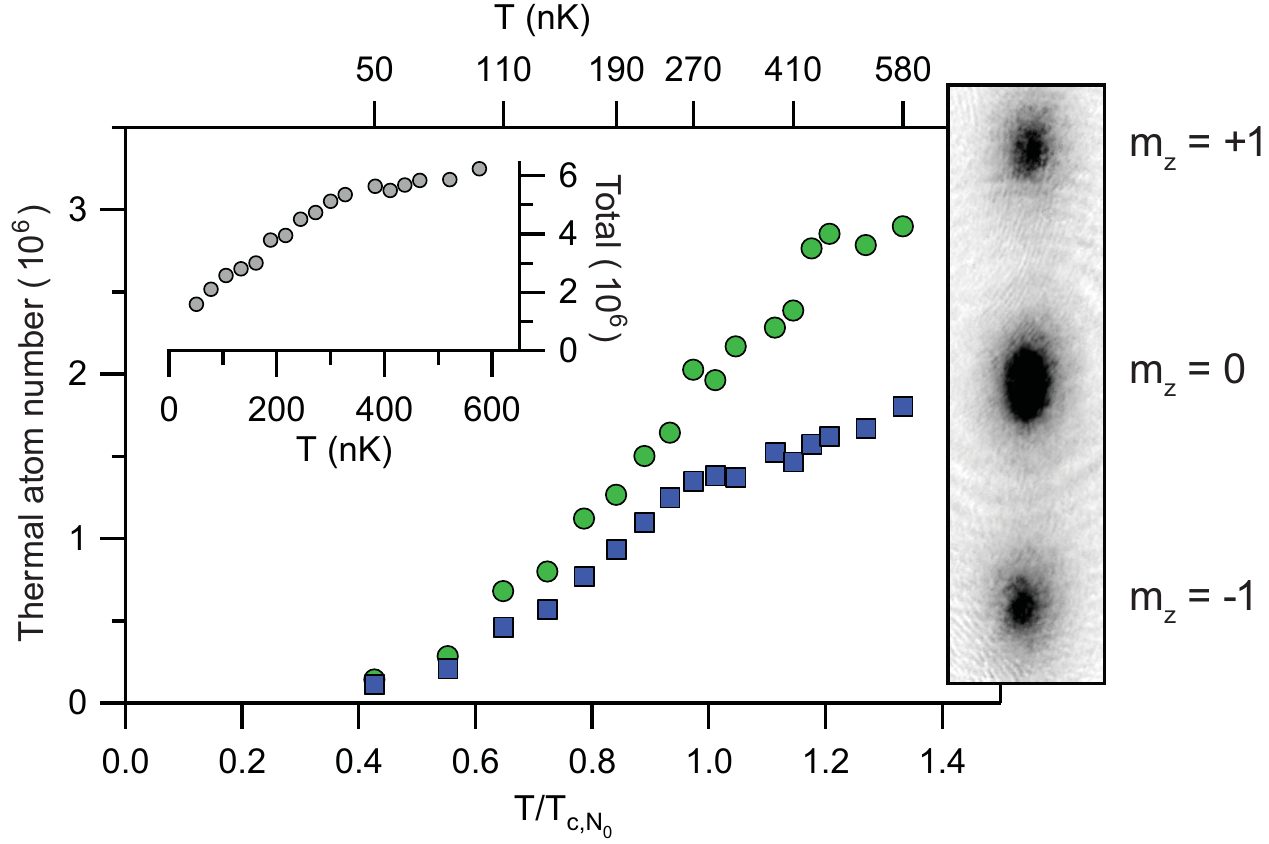}
\caption{The number of atoms in the thermal fraction of the
$|m_z=0\rangle$ sublevel (green circles) is compared with the
average number in the thermal fractions of the $|m_z=\pm 1\rangle$
states (blue squares) at different final temperatures of the $\eta =
1/4$ clouds. These data are derived from bimodal fits to
time-of-flight density images taken after the three spin states were
spatially separated in a magnetic field gradient (such an image is
shown at right).  While the relative populations in the Zeeman
levels at high temperature are consistent with the initial value of
$\eta = 1/4$, below $T_{c,N_0}$, the thermal components in all
Zeeman components become roughly equal.  A record of the total atom
number (condensed and non-condensed) in all three sublevels vs.\
temperature is shown in the inset.} \label{fig:number_figure}
\end{figure}

After the optical trap depth was reduced to the desired level, the
spinor gas was allowed to equilibrate, typically for another 200 ms,
and was then probed by either of two methods. The instantaneous
vector magnetization $\tilde{\bf M} = \mu \tilde{n} {\bf F}$,
column-integrated along the $\hat{y}$ imaging axis, was measured in
the $\hat{x}$-$\hat{z}$ plane by magnetization-sensitive phase
contrast imaging, at a measured imaging resolution of about 2 $\mu$m
\cite{higb05larmor,veng07mag}. Here, $\mu$ is the atomic magnetic
moment and $\tilde{n}$ is the local column density of the gas.
Alternately, to measure the temperature, atom number and condensate
fraction for each of the Zeeman sublevels, the gas was released from
the optical trap and subjected to a magnetic field gradient that
served to separate spatially the three Zeeman states. The three
components were then imaged following an additional time of flight
(Fig.\ \ref{fig:number_figure}).

Throughout the evaporation and equilibration, a  $B = 150$ mG
magnetic field, varying by less than 5 $\mu$G across the extent of
the gas, was applied along $\hat{z}$. This field produces a
quadratic Zeeman shift of the form $q F_z^2$ with $q/h = (70 \,
\mbox{Hz}/\mbox{G}^2) B^2 = 1.5$ Hz.  This shift is smaller than the
spin-dependent contact and dipolar interactions in the degenerate
gases studied, with typical energies of $|c_2| n = h\times 7$ Hz and
$\mu_0 \mu^2 n = h \times 8$ Hz, respectively, at a typical density
of $n = 2 \times 10^{14} \, \mbox{cm}^{-3}$. The field also induces
rapid (110 kHz) Larmor precession of the atomic spins, owing to
which the magnetic dipole interactions assume a precession-averaged
form \cite{kawa07observe,veng08helix,cher09roton}.

\section{Magnetization textures produced from $\eta = 0$ gases}

Spinor gases produced initially with $\eta = 0$ correspond to the
equilibrium state far from quantum degeneracy for our system where
$k_B T \gg \{q, |c_2| n\}$. The bulk features of gases produced by
cooling this initial mixture display several hallmarks of Bose
condensation. Gases probed after a time of flight show the
transition from a Bose-enhanced gaussian distribution to a bimodal
density distribution at a temperature consistent with the ideal-gas
Bose-Einstein condensation temperature, $T_{c,N_0}$, given as  $k_B
T_{c, N_0} = \hbar \bar{\omega} (N_0/1.21)^{1/3}$ where
$\bar{\omega} = (\omega_x \omega_y \omega_z)^{1/3}$ and $N_0$ is the
atom number in the $|m_z = 0\rangle$ state.  The measured
populations within the central peak and the gaussian distribution,
associated typically with the condensate and thermal fractions,
respectively, match closely with those expected for scalar Bose
gases.  The total population in each Zeeman sublevel remained
roughly equal as the gas was cooled.

\begin{figure}[tb]
\centering
\includegraphics[width=0.45 \textwidth]{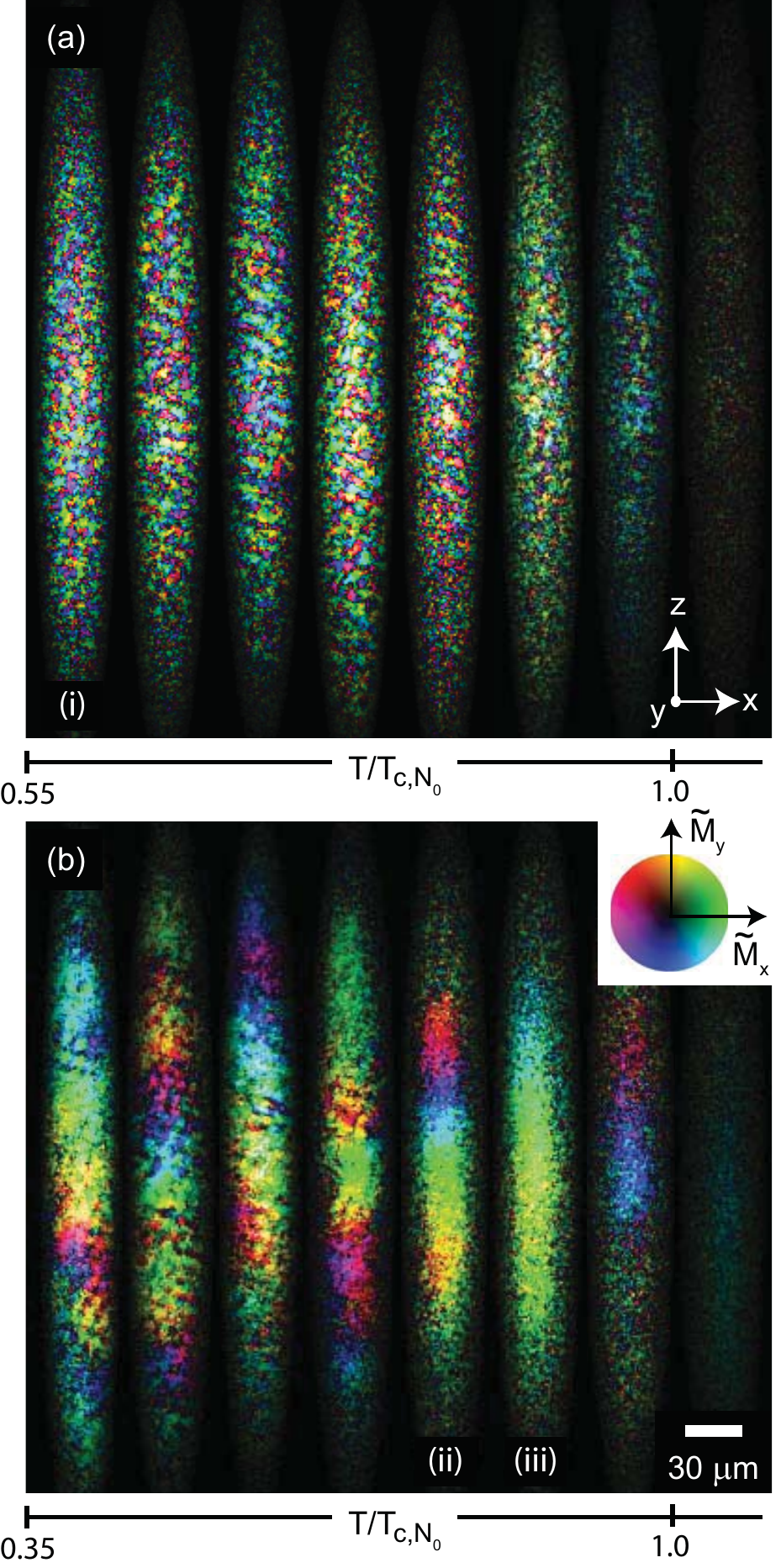}
\caption{Spin textures of $^{87}$Rb $F=1$ spinor gases produced by
cooling thermal spin mixtures with initial values (a) $\eta = 0$ and
(b) $\eta = 1/4$. The transverse magnetization $\tilde{M}_{x,y}$
(color wheel shown) is shown in the $\hat{x}$-$\hat{z}$ plane.
Temperatures are scaled by the Bose condensation temperature given
the trap parameters and the population of the $|m_z = 0\rangle$
state.  Textures labeled (i), (ii) and (iii) are analyzed further in
Fig.\ \ref{fig:fourier}.} \label{fig:vstemp}
\end{figure}

\begin{figure}[tb]
\centering
\includegraphics[width=0.4 \textwidth]{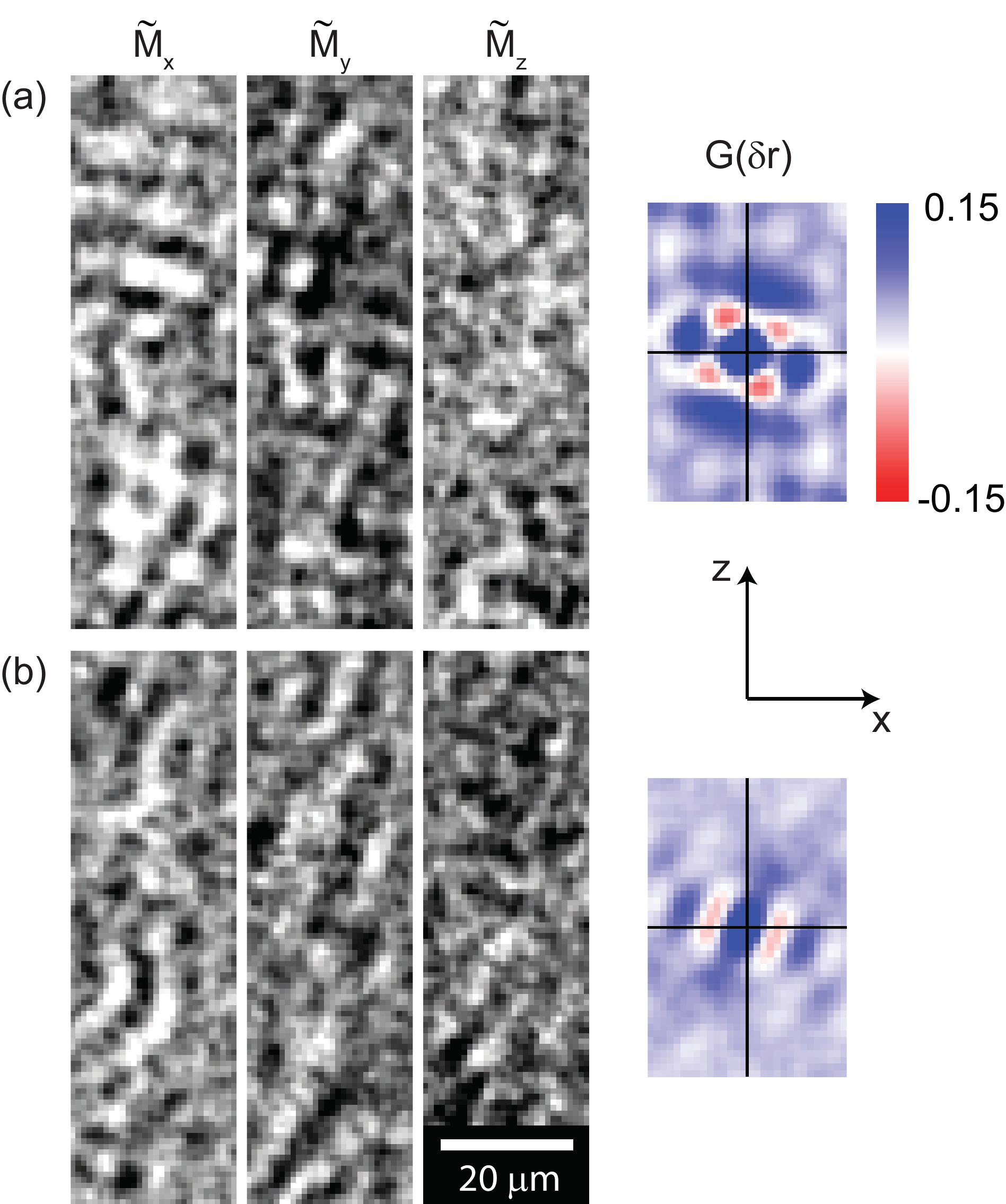}
\caption{The magnetization vector components $\tilde{M}_{x,y,z}$
(grayscale spanning $\pm 2/3$ of the maximum condensate
magnetization) are shown for the central $25 \times 100 \,
\mu\mbox{m}^2$ regions of two samples produced by cooling initial
mixtures with $\eta = 0$ to 80 nK. Regions of sample (a) show
prominent magnetization modulations along either a single
near-vertical wavevector (top portion) or along two orthogonal
wavevectors (bottom portion).  In contrast, modulations in (b) are
predominantly along a near-horizonal wavevector.  The magnetization
spatial correlation function $G(\delta r)$ evaluated over a $40
\times 200 \, \mu\mbox{m}^2$ region reveals  patterns of alternating
spins characteristic of these samples.} \label{fig:samples}
\end{figure}

However, {\em in situ} probing of these gases (Fig.\
\ref{fig:vstemp}) reveals the formation of complex magnetic
structure. Immediately below $T_{c,N_0}$, the spinor gas becomes
spontaneously magnetized, breaking spin-rotational symmetry within a
central region consistent with the spatial dimension of a Bose
condensate. The magnetization observed for two samples produced
under similar conditions is shown in Fig.\ \ref{fig:samples}. These
inhomogeneous textures are dominated by strong modulation of the
magnetization with a characteristic domain size $l$ of about $5
\,\mu$m.  We observed spin domains of similar size even in gases
with reduced atom number for which the condensate radii were roughly
half those shown in Fig.\ \ref{fig:vstemp}.  That this length $l$ is
smaller than the condensate dimensions in the imaged
$\hat{x}$-$\hat{z}$ plane and is not strongly dependent on the
$\hat{x}$ dimension of the condensate suggests that the observed
translational symmetry breaking is an intrinsic tendency of this
quantum gas.

We note also that both $l$ and also the spin healing length defined
as $\xi_s = \hbar/(2 m |c_2| n)^{1/2} = 3 \, \mu$m exceed the
Thomas-Fermi condensate radius along the direction of tightest
confinement, $r_y \simeq 1.5 \, \mu$m. Thus, spin textures of the
degenerate gas may be considered two-dimensional.  In contrast, the
scalar healing length, defined conventionally as $\xi = \hbar / (2 m
c_0 n)^{1/2} = 0.2 \,\mu\mbox{m}$ with $c_0$ defining the
spin-independent contact interaction strength, is smaller than the
Thomas-Fermi radii of the condensate.  Thus, in terms of scalar
excitations, the degenerate gas may be considered three-dimensional.

\begin{figure}[tb]
\centering
\includegraphics[width=0.4 \textwidth]{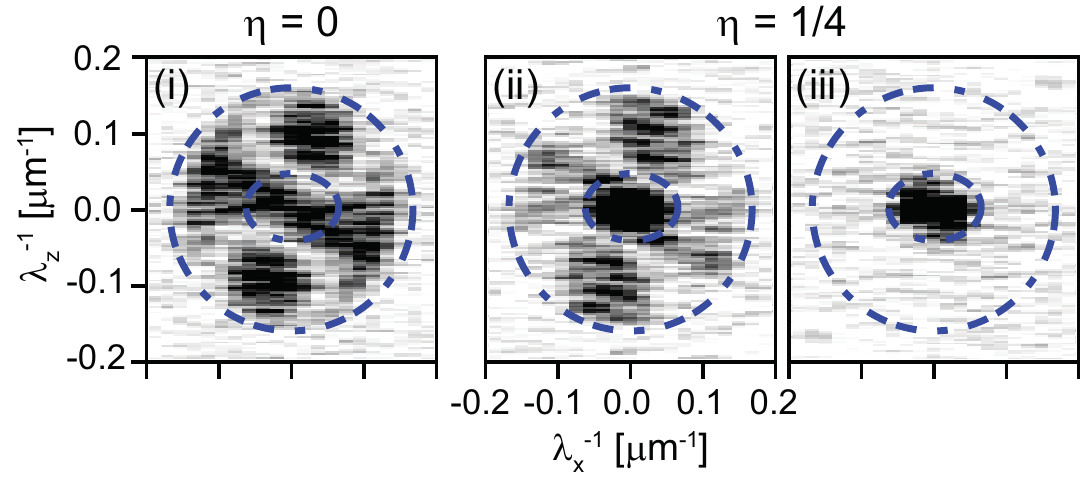}
\caption{Spatial Fourier power spectra of the magnetization,
$|\tilde{\bf{M}}(k_x, k_y)|^2$ for gases prepared with $\eta=0$ (i)
and $\eta = 1/4$ (ii, iii). Periodic spin modulation in (i) and (ii)
is indicated by broad spectral peaks with modulation wavelength
$\lambda \simeq 10 \,\mu$m. Ferromagnetic order in (ii) and (iii) is
indicated by the spectral peak at zero wavevector. (ii) and (iii)
correspond to gases cooled to temperatures of 100 nK and 130 nK
respectively, indicating the abrupt transition to a modulated spin
texture.  The power spectrum is evaluated for the entire
magnetization texture, corresponding to images shown in Fig.\
\ref{fig:vstemp}.    Dashed ellipses show demarcation of Fourier
space into the central region, indicative of ferromagnetic order,
and an annular region, indicative of short-range spin modulation.
Such regions are used for an empirical characterization (Fig.\
\ref{fig:spinmeasures}) of the spin textures. \label{fig:fourier}}
\end{figure}

Importantly, these textures display spatial order, characterized by
strong modulation along two distinct wavevectors.  To illustrate
this spatial order, we consider the spatial correlation function of
the vector magnetization, defined as
\begin{equation}
G(\delta {\bf r}) = \frac{\sum_{\bf r} \tilde{\bf M}({\bf r} +
\delta {\bf r}) \cdot \tilde{\bf M}({\bf r})}{\mu^2 \sum_{\bf r}
\tilde{n}({\bf r} + \delta {\bf r})  \tilde{n}({\bf r})}
\end{equation}
and shown in Fig.\ \ref{fig:samples}.  The characteristic spatial
pattern is indicated by the lobes of positive and negative spin
correlation surrounding the central region of positive correlation
at $\delta \bf{r} = 0$.  This spatial organization is equivalently
indicated by peaks in the Fourier power spectrum of the vector
magnetization, $|\tilde{{\bf M}}(k_x, k_z)|^2$, at wavevectors $k
\simeq \pi/l$ (Fig.\ \ref{fig:fourier}) \cite{note:pixelation}.  We
highlight three characteristics of this spatial order.  First, the
alternating domain pattern is short-ranged, extending regularly only
over a typical range of 30 $\mu$m that is much shorter than the
maximum extent ($\simeq 300\,\mu$m at low temperatures) of the
magnetized degenerate gas. Second, the modulation pattern varies
visibly between samples produced at similar temperatures and
equilibration times, with the spin texture in some regions
characterized by a checkerboard, or two-wavevector pattern of
magnetization, while in other regions the modulation along one
wavevector appears more dominant. In spite of these variations, the
orientation of the two modulation wavevectors is fairly constant
across samples. We observe that a rotation of the optical trap about
$\hat{y}$ causes a corresponding rotation in the crystal
wavevectors, demonstrating that the orientation of the
magneto-crystalline pattern in the $\hat{x}$-$\hat{z}$ plane is
pinned to the boundaries imposed by the trap.  We cannot account for
a misalignment of these wavevectors from the apparent trap axes,
though we suspect it arises from trap asymmetries due to aberrations
in the optical trapping laser beam.

\begin{figure}[b]
\centering
\includegraphics[width=0.49\textwidth]{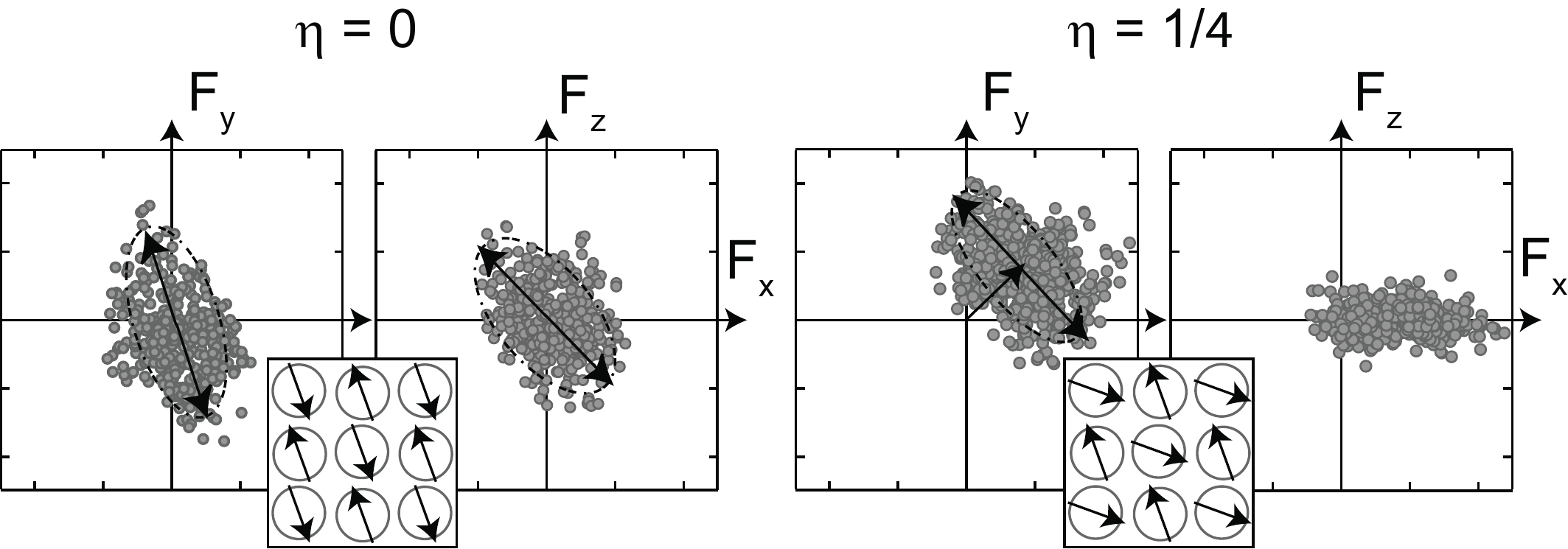}
\caption{The periodic magnetization textures are characterized by a
spin-space axis along which the magnetization is modulated.
Distributions of spins within the central $30 \times 20 \,
\mu\mbox{m}^2$ regions of the magnetization texture, for samples at
the lowest temperatures. For $\eta = 0$, both transverse ($F_{x,y}$)
and longitudinal ($F_z$) spin variations are evident, while for
$\eta = 1/4$ the modulation is solely transverse. The long axis of
the prolate spin distribution defines the local spin axis (indicated
by double-headed arrow).  For $\eta = 1/4$, this distribution is
offset due to the coexistent ferromagnetic order. Spatial patterns
corresponding to the observed spin-space modulation are shown
schematically at bottom.  Dashed ellipses are guides to the eye.}
\label{fig:spinaxis}
\end{figure}

\begin{figure}[tb]
\centering
\includegraphics[width=0.35\textwidth]{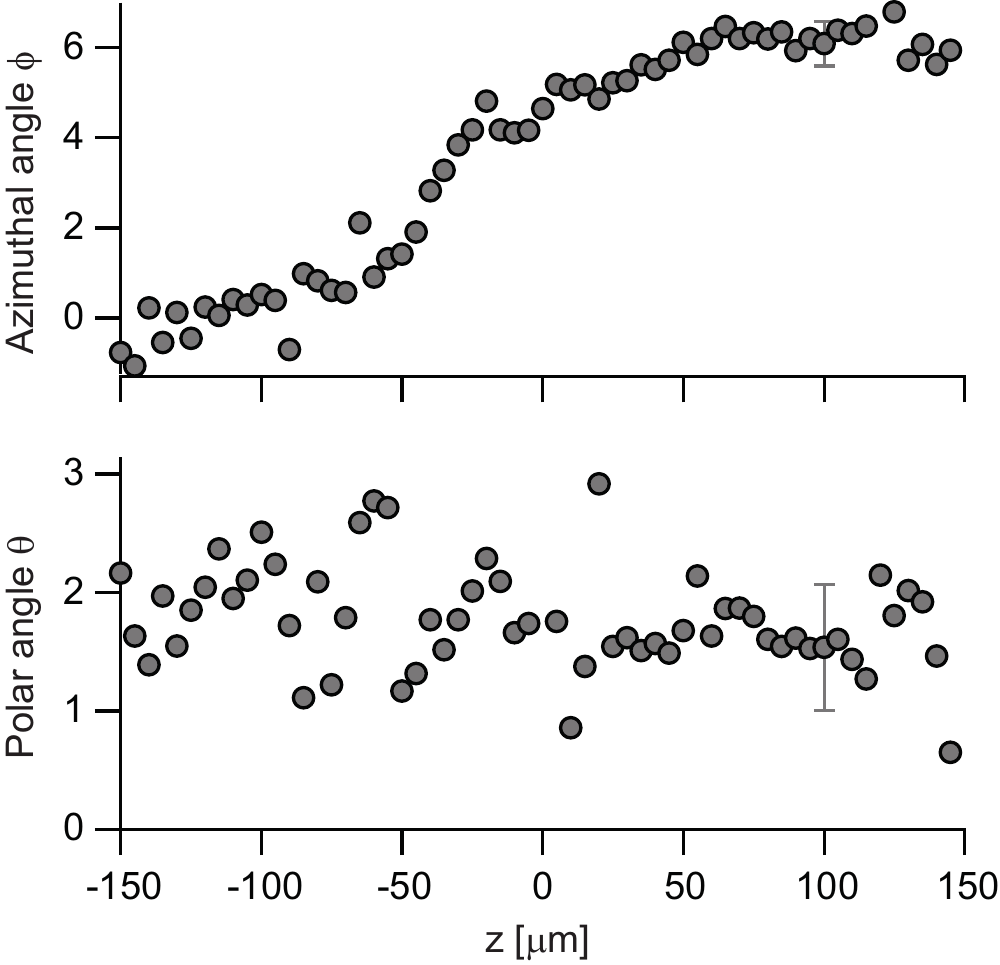}
\caption{The local spin axis, determined with a $30 \time 20 \,
\mu$m region centered variably along the $\hat{z}$ axis, is not
uniform across the extent of the condensate but meanders gradually
over length scales of around 50 $\mu$m. The azimuthal and polar
angles of this axis, given in radians, with typical statistical
error bar shown, are presented for the lowest-temperature sample in
Fig.\ \ref{fig:vstemp}b, with initial $\eta = 1/4$.}
\label{fig:axisvsposition}
\end{figure}

Third, the magnetization modulation of these spin textures is
characterized by an axis in spin space along which the magnetization
exhibits the largest variance.  To identify this local spin axis, we
considered the distribution of vector spins  $\vec{F} =
(F_x,F_y,F_z)$ measured at each imaged pixel within $30 \times 20 \,
\mu\mbox{m}^2$ regions of the gas (Fig.\ \ref{fig:spinaxis}).  These
regions were centered on the condensate in the $\hat{x}$ direction
while the location along $\hat{z}$ was allowed to vary.  The
observed spin-space distribution was significantly prolate
\cite{note:significance}. The local spin axis was then defined as
lying along the largest-eigenvalue eigenvector of the covariance
matrix $D_{ij} = \langle (F_i - \langle F_i\rangle)(F_j - \langle
F_j \rangle)\rangle$ with $\{i,j\} \in \{x,y,z\}$.  This spin axis
varies over characteristic distances of $\simeq$ 50 $\mu$m across
the length of the spin texture (Fig.\ \ref{fig:axisvsposition}).

\section{Magnetization textures produced from $\eta \neq 0$ gases}

We also studied the evolution of spin mixtures prepared with a
non-zero initial quadrupole moment ($\eta = 1/4$). While the small
rate of spin-mixing collisions in the non-degenerate gas prevented
its equilibration to $\eta=0$ within experimentally accessible
timescales (seconds), below $T_{c,N_0}$, the thermal fraction of
degenerate gases did reach roughly equal populations in the three
Zeeman sublevels (Fig.\ \ref{fig:number_figure}), presumably due to
the high density and bosonic enhancement intrinsic to a condensate.
These thermal populations were all consistent with the expected
non-condensate population for a quantum-degenerate, thermally
equilibrated, harmonically confined Bose gas.

Yet, in spite of this condensate-mediated spin mixing, effects of
the initial condition were still visible in the magnetization
textures of the degenerate gas.  In contrast to the case of $\eta =
0$, here we observe two distinct magnetization patterns. Just below
the condensation temperature, the spin textures showed ferromagnetic
order, with a nearly uniform magnetization that varied only at long
length-scales of around 100 $\mu$m.  The stability of these
transverse ferromagnetic textures shows that the presence of a
condensate fraction in the $|m_z = \pm 1\rangle$ Zeeman sublevels
states does not on its own guarantee the appearance of small spin
domains.  Only at a distinctly lower temperature does the
magnetization display periodic spatial modulations.   This
modulation pattern coexists with ferromagnetic order, constituting a
spatial variation of the magnetization about a non-zero average
value.

The distinction between long length-scale ferromagnetic order and
the short length-scale spin modulation is clearly seen in the
spatial Fourier power spectra of the magnetization (Fig.\
\ref{fig:fourier}). This distinction allows us to define empirical
measures of each type of spatial order.  For this, we demarcate
spatial Fourier space into a central region, which we may associate
with ferromagnetic order, surrounded by an annular region which
contains the broad spectral peaks that characterize the short
length-scale spin modulation. Summing the spectral power in each of
these regions provides a measure of the different magnetization
patterns exhibited by the quantum gas.

\begin{figure}[tb]
\centering
\includegraphics[width=0.4 \textwidth]{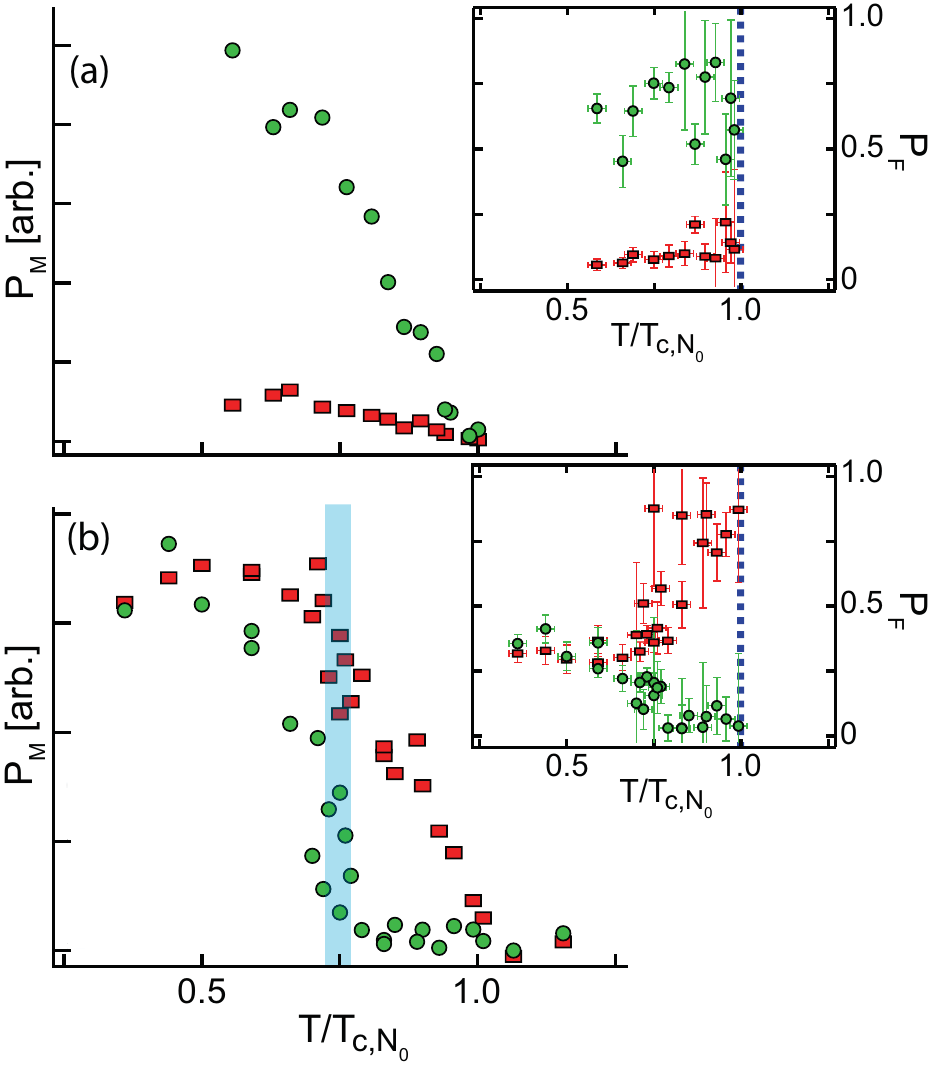}
\caption{Measures for ferromagnetic order (red rectangles) and for
the short length-scale spin modulation (green circles) are defined
as the sums $P_M = \sum_k |\tilde{\bf M}(k_x, k_y)|^2$ in regions of
spatial Fourier space defined in Fig.\ \ref{fig:fourier}. Insets
show these order parameters as fractions of the integrated spectral
power expected for a uniformly magnetized condensate.  (a) Gases
cooled from $\eta = 0$ show a single transition to a magnetic
texture dominated by short-range spin modulation. (b) Data for $\eta
= 1/4$ show an additional transition between textures with
coexistent ferromagnetic order and spin modulation observed at low
temperature, and solely ferromagnetic textures observed at high
temperature.} \label{fig:spinmeasures}
\end{figure}

These measures distinguish between spin textures obtained from
differing initial conditions (Fig.\ \ref{fig:spinmeasures}). For
gases produced with $\eta = 0$, the modulated spin pattern arises
immediately below $T_{c,N_0}$ and intensifies monotonically with
decreasing temperature, remaining a constant fraction of the
spectral power expected for a fully magnetized condensate. The
ferromagnetic order remains small at all temperatures indicating
that the average vector magnetization over length scales larger than
$\sim 20 \, \mu$m is roughly zero.  In contrast, for $\eta = 1/4$,
ferromagnetic order dominates down to $0.75 \, T_{c,N_0} =$ 130 nK.
Below this temperature, the short length-scale spin modulation
emerges abruptly, as indicated by the different Fourier spectra
shown in Fig.\ \ref{fig:fourier} (ii) and (iii), and coexists with
ferromagnetism.  Here we note that the magnetic dipole energy
between neighboring spin domains may be approximated as $\mu_0 \mu^2
\tilde{n}^2 l / 4 \pi = k_B \times 210$ nK. Thus, dipolar
interactions may play a role in the transition toward spatially
modulated spin textures.

In addition to the presence of the two distinct magnetization
patterns, the textures produced from initial conditions $\eta =0$
and $\eta=1/4$ also differ in the overall distribution of
magnetization orientations. For $\eta = 0$, the magnetization
orientation shows no preferred axis or plane, while for $\eta = 1/4$
the magnetization was found to lie predominantly in the (transverse)
$\hat{x}$-$\hat{y}$ plane.

This spin-space asymmetry may be explained in part by the fact that,
for $\eta > 0$, Bose condensation occurs first in the dominant $|m_z
= 0\rangle$ state. Transverse magnetization may then arise from the
dynamical instability of this polar-state condensate
\cite{sadl06symm}.  This timeline for the formation of magnetization
was verified by probing gases at different times during their
evaporation and equilibration. Indeed, with $\eta = 1/4$,
unmagnetized condensates in the $|m_z = 0\rangle$ state were
observed before the appearance of magnetization. Further, at $q = h
\times 70$ Hz $ > 2 |c_2| n$, for which the quadratic shift
dominates over both the spin-dependent collisional and dipolar
interactions and, thus, a $|m_z = 0\rangle$ condensate is stable, a
similar cooling sequence resulted neither in a condensate fraction
in the $|m_z = \pm 1\rangle$ states nor in the appearance of
transverse magnetization.

\section{Implications for equilibrium phases}

The limited lifetime (about 1 s) of our gaseous samples restricted
the duration over which they were allowed to equilibrate. This
restriction raises the question as to what extent the observed
magnetic textures display the characteristics of thermal equilibrium
phases.  To help address this question, we have verified that both
the temperature and the populations of the non-condensate fractions
reach equilibrium within 100 - 150 ms of the establishment of the
final trap depth. Doubling the evaporation time or lengthening the
equilibration period to as long as 800 ms produced no discernible
changes in the magnetic order, other than overall loss in atom
number. Nevertheless, as discussed above, we observe variations in
the magnetization patterns in samples produced under similar
conditions, and persistent differences between samples prepared from
differing initial spin mixtures.  These variations suggest that the
magnetization of these degenerate spinor gases equilibrates only
slowly, a property observed also in classical dipolar systems
\cite{lott91}.  Therefore, the observed textures and the empirical
measures used to characterize these textures cannot be considered to
define the equilibrium phase of the degenerate spinor gas.

However, the presence of short-range periodic spin modulation is a
robust property of the observed spin textures, appearing
consistently under a variety of initial conditions.  We have
additionally observed similar modulation patterns following the
dissolution of helical spin textures \cite{veng08helix} as well as
after long evolution times following a quench of an unmagnetized
condensate to a magnetic state (at short evolution times, this
instability produces a distinctly different modulation pattern)
\cite{sadl06symm,lesl09amp}. This robustness, and the fact that the
characteristic length-scale of the spin modulation is constant and
much smaller than the extent of the gaseous sample (certainly in the
$\hat{z}$ direction), supports the claim that such structure is a
characteristic of the low-temperature equilibrium configuration of
the bulk quasi-two-dimensional spinor Bose gas.

We thank S.M.\ Girvin, T.L.\ Ho, J.\ Moore and A.\ Vishwanath for
valuable discussions. This work was supported by the NSF, and by the
DARPA OLE program (ARO Grant No.\ W911NF-07-1-0576). Partial
personnel and equipment support was provided by the Division of
Materials Sciences and Engineering, Office of Basic Energy Sciences.
S.R.L.\ acknowledges support from NSERC and F.S.\ from the Gottlieb
Daimler and Karl Benz Foundation.


\begin{thebibliography}{10}
\bibitem{note:pixelation}%
  \BibitemOpen
  \bibinfo {note} {$G(\delta\mathbf{r})$ is accurate only over distances
  substantially smaller than the extent of the finite field over which it is
  evaluated. The correlation functions presented in Fig.\ \ref{fig:samples}
  satisfy that criterion. Equivalently, the Fourier power spectrum is pixilated
  due to the finite spatial extent of the data field, and accurate measures can
  only be obtained for spectral features that span many pixels in Fourier
  space. The features shown in Fig.\ \ref{fig:fourier} are all broad, owing to
  the short-range nature of the periodic spin modulation as shown by the
  correlation function and in the data themselves.}%
  \bibAnnoteFile{Stop}{note:pixelation}%
\bibitem{note:significance}%
  \BibitemOpen
  \bibinfo {note} {Based on numerical simulations that take into account the
  signal to noise ratio of the magnetization-sensitive images, the number of
  independent measurements within the region of interest, and assuming a
  gaussian noise distribution, we determined a cutoff of 1.35 for the aspect
  ratio of the ellipsoidal distribution (major axis radius divided by geometric
  mean radius) below which the spin axis was no longer well defined. This
  cutoff was also verified by examining the spin distributions in homogenously
  magnetized condensates which did not exhibit short-range spin modulations.}%
  \bibAnnoteFile{Stop}{note:significance}%
\end{thebibliography}

%

\end{document}